# *Herschel* SPIRE FTS Relative Spectral Response Calibration

**Trevor Fulton • Rosalind Hopwood •**
**Jean-Paul Baluteau • Dominique Benielli •**
**Peter Imhof • Tanya Lim •**
**Nanyao Lu • Nicola Marchili •**
**David Naylor • Edward Polehampton •**
**Bruce Swinyard • Ivan Valtchanov**

T. Fulton
Bluesky Spectroscopy, Lethbridge, AB, T1J 0N9, Canada
Institute for Space Imaging Science, Department of Physics & Astronomy, University of Lethbridge, Lethbridge, AB, T1K3M4, Canada

R. Hopwood
Physics Department, Imperial College London, South Kensington Campus, SW7 2AZ, UK

J.-P. Baluteau
Aix Marseille Université, CNRS, LAM (Laboratoire d'Astrophysique de Marseille) UMR 7326, 13388, Marseille, France

D. Benielli
Aix Marseille Université, CNRS, LAM (Laboratoire d'Astrophysique de Marseille) UMR 7326, 13388, Marseille, France

P. Imhof
Bluesky Spectroscopy, Lethbridge, AB, T1J 0N9, Canada
Institute for Space Imaging Science, Department of Physics & Astronomy, University of Lethbridge, Lethbridge, AB, T1K3M4, Canada

T. Lim
RAL Space, Rutherford Appleton Laboratory, Didcot OX11 0QX, UK

N. Lu
NHSC/IPAC, 100-22 Caltech, Pasadena, CA 91125, USA

N. Marchili
Universitá di Padova, I-35131 Padova, Italy

D. Naylor
Institute for Space Imaging Science, Department of Physics & Astronomy, University of Lethbridge, Lethbridge, AB, T1K3M4, Canada

E. Polehampton
RAL Space, Rutherford Appleton Laboratory, Didcot OX11 0QX, UK;
Institute for Space Imaging Science, Department of Physics & Astronomy, University of Lethbridge, Lethbridge, AB, T1K3M4, Canada

B. Swinyard

RAL Space, Rutherford Appleton Laboratory, Didcot OX11 0QX, UK
Dept. of Physics & Astronomy, University College London, Gower St, London, WC1E 6BT, UK

I. Valtchanov
Herschel Science Centre, ESAC, P.O. Box 78, 28691 Villanueva de la Cañada, Madrid, Spain

**Abstract**

*Herschel*/SPIRE Fourier transform spectrometer (FTS) observations contain emission from both the *Herschel* Telescope and the SPIRE Instrument itself, both of which are typically orders of magnitude greater than the emission from the astronomical source, and must be removed in order to recover the source spectrum. The effects of the *Herschel* Telescope and the SPIRE Instrument are removed during data reduction using relative spectral response calibration curves and emission models. We present the evolution of the methods used to derive the relative spectral response calibration curves for the SPIRE FTS. The relationship between the calibration curves and the ultimate sensitivity of calibrated SPIRE FTS data is discussed and the results from the derivation methods are compared. These comparisons show that the latest derivation methods result in calibration curves that impart a factor of between 2 and 100 less noise to the overall error budget, which results in calibrated spectra for individual observations whose noise is reduced by a factor of 2-3, with a gain in the overall spectral sensitivity of 23% and 21% for the two detector bands, respectively.

**Keywords** SPIRE • FTS • Spectroscopy • Calibration • Sensitivity

## 1. INTRODUCTION

The Spectral and Photometric Imaging REceiver (SPIRE) (Griffin et al, 2010) was one of three focal plane instruments aboard the ESA Herschel Space Observatory (Pilbratt et al, 2010). SPIRE consisted of an imaging photometric camera and an imaging Fourier transform spectrometer (FTS). The SPIRE FTS provided simultaneous wide frequency coverage in the sub-millimetre using two hexagonally packed bolometer arrays: SLW (447-990 GHz) and SSW (958-1546 GHz).

The Mach-Zehnder design of the SPIRE FTS (Ade et al 1999, Dohlen et al 2000, Swinyard et al 2010), shown in Fig. 1, means that the radiation incident on the two detector arrays is a combination of that from the two input ports; the telescope port, and the instrument port (Swinyard et al, 2013). Each SPIRE detector measures the interference pattern obtained by scanning the spectrometer mirror mechanism (SMEC), which is then converted, via a Fourier transform, to a spectrum in units of voltage density in V GHz$^{-1}$ (Fulton et al, 2013). The measured voltage density spectra, $V_{Meas}(\nu)$, can be expressed as a linear combination of spectral contributions from three distinct entities: the astronomical source, $V_{Source}(\nu)$ via the telescope port; the *Herschel* Telescope, $V_{Tel}(\nu)$ via the telescope port; and the SPIRE Instrument, $V_{Inst}(\nu)$ via the instrument port (Fulton et al, 2013),

$$V_{Meas}(\nu) = V_{Source}(\nu) + V_{Tel}(\nu) + V_{Inst}(\nu). \qquad (1.1)$$

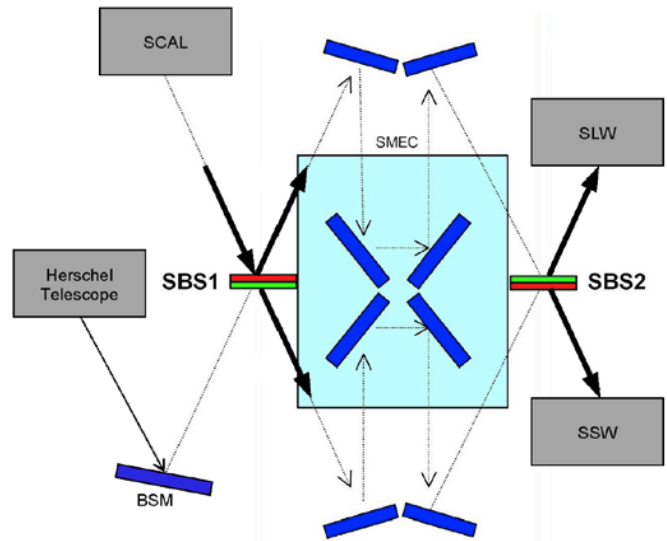

**Fig. 1: Simplified layout of the SPIRE Spectrometer, showing the two input ports from the Herschel Telescope and the SPIRE Instrument (SCAL), the Beam Steering Mirror (BSM), two beam splitters (SBS1 and SBS2), Spectrometer mirror mechanism (SMEC) and two detector arrays (SLW and SSW).**

The instrument port is terminated on a cold internal radiative source, SCAL (Hargrave et al, 2006), which was initially intended to "null" the emission of the *Herschel* Telescope in order to increase the dynamic range of the SPIRE FTS detectors (Swinyard et al, 2013). Results of tests performed during the Performance Verification (PV) phase of the mission deemed such a "nulling" source unnecessary (Swinyard et al, 2010), and so SCAL was not actively heated during the operation phase of the mission. The SCAL emitter was therefore in thermal equilibrium with the SPIRE Focal Plane Unit, with a temperature range of 4.5 K to 5.2 K over the course of the mission.

The SPIRE FTS offered low spectral resolution (~10 GHz) and high spectral resolution (1.185 GHz) in either sparse spatial sampling (2 beam spacing) or mapping (1 or ½ beam spacing) modes. The high resolution, sparse mode was used for the majority of SPIRE observations – 68% of the total time that SPIRE was observing – therefore, we concentrate only on the high resolution mode in this paper.

In order to calibrate data from the SPIRE FTS, the contributions to the final spectrum from both input ports must be carefully considered. In this paper, we describe the emission models adopted to derive the Relative Spectral Response Functions (RSRFs) applicable to each port in Section 2. Section 3 and Section 4 describe the calculation of the instrument and telescope RSRFs, respectively, including the improvements to the methods that were developed for the Herschel Interactive Processing Environment (HIPE; Ott et al, 2010) v11. Section 5 compares the different methods, and quantifies the improvements made for HIPE v11.

For the purposes of this paper, the RSRF is defined as the ratio of the measured voltage density spectrum in V GHz$^{-1}$ and the intensity spectrum in units of W m$^{-2}$ Hz$^{-1}$ sr$^{-1}$.

## 2. SPIRE EMISSION MODELS

The housekeeping telemetry from the *Herschel* Telescope indicates that although the telescope temperature was stable to ± 10 mK over the course of a single SPIRE FTS observation, it varied by ±2K over the course of the mission (Hopwood et al, 2013). This variation translates to a difference in flux density of ±5 Jy and ±10 Jy at the low and high frequency ends of the SLW band, and ±3-5 Jy in the SSW band. In addition, the SPIRE instrument temperature varied between 4.5K and 5.2K over the course of the mission, which translates to a variation in flux density on the order of ±10 Jy at the low frequency end of SLW. This shows that the background conditions of both telescope and instrument varied significantly throughout the mission.

In order to compensate for the changing background conditions of the SPIRE FTS, a calibration plan based on dynamic emission models of the two input ports was developed. As a starting point, a simplified version of the SPIRE FTS Instrument shown in Fig. 1 was adopted.

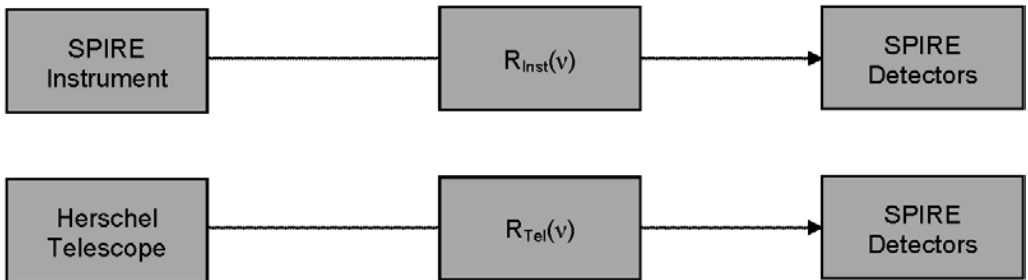

**Fig. 2**: **Simplified SPIRE Emission Model.**

In this simplified model (Fig. 2), the voltage density spectra from the SPIRE Instrument and the *Herschel* Telescope, are each given by the product of a blackbody model, $M_{Inst}(T_{Inst}, \nu)$ and $M_{Tel}(T_{Tel}, \nu)$, and the RSRF for the emission path, $R_{Inst}(\nu)$ and $R_{Tel}(\nu)$, as

$$V_{Inst}(\nu) = M_{Inst}(T_{Inst}, \nu)R_{Inst}(\nu) \tag{2.1}$$
$$V_{Tel}(\nu) = M_{Tel}(T_{Tel}, \nu)R_{Tel}(\nu) \tag{2.2}$$

The model of the SPIRE Instrument is the Planck function,

$$M_{Inst}(T_{Inst}, \nu) = B(T_{Inst}, \nu) = \frac{2h\nu^3}{c^2} \frac{1}{e^{\frac{h\nu}{kT_{Inst}}} - 1} \qquad [\text{W m}^{-2} \text{ Hz}^{-1} \text{ sr}^{-1}] \qquad (2.3)$$

where $T_{Inst}$ is the average of the temperature of the SPIRE Instrument over the course of a given scan of the SMEC. This temperature is measured by a thermometer located on the SCAL source that fills the instrument port.

The model of the *Herschel* Telescope, $M_{Tel}(T_{Tel}, \nu)$, is expressed as

$$M_{Tel}(T_{Tel}, \nu) = (1 - \varepsilon_{M2})E_{corr}\varepsilon_{M1}B(T_{M1}, \nu) + \varepsilon_{M2}B(T_{M2}, \nu) \quad [\text{W m}^{-2} \text{ Hz}^{-1} \text{ sr}^{-1}]. \qquad (2.4)$$

where $\varepsilon$ is the emissivity of mirrors M1 and M2 (Fischer et al, 2004), $T_{M1}$ and $T_{M2}$ ($T_{Tel}$) are the average temperatures of mirrors M1 and M2, and $E_{corr}$ is a time dependent correction to $\varepsilon_{M1}$ as described in detail in Hopwood et al (2013).

The overall measured voltage density spectrum, taking into account the emission models, for any SPIRE FTS observation is therefore

$$V_{Meas}(\nu) = V_{Source}(\nu) + M_{Tel}(T_{Tel}, \nu)R_{Tel}(\nu) + M_{Inst}(T_{Inst}, \nu)R_{Inst}(\nu). \qquad (2.5)$$

To obtain the spectrum of the telescope and instrument independently of any astronomical source, we observed a patch of "dark sky" towards the ecliptic pole at RA:17h40m12s, Dec:+69d00m00s (J2000). Disregarding negligible contributions from the cosmic background, the dark sky spectrum can be expressed as

$$V_{Dark}(\nu) = M_{Tel}(T_{Tel}, \nu)R_{Tel}(\nu) + M_{Inst}(T_{Inst}, \nu)R_{Inst}(\nu). \qquad (2.6)$$

## 3. INSTRUMENT CALIBRATION

In order to derive the value of the instrument RSRF, $R_{Inst}(\nu)$, observations of dark sky were used with a scan-by-scan differencing method. This allowed $R_{Inst}(\nu)$ to be isolated from the contribution of the telescope.

The plot in Fig. 3 shows a strong variation in temperature of the SPIRE Instrument, represented by the SCAL thermometer, over the course of an observation of dark sky with 80 SMEC scans. The oscillations seen in the temperature correspond to the individual scans of the SMEC during the observation. Each SMEC scan produces a spectrum, and the voltage density of the $n^{th}$ spectrum can be represented as

$$V_{Dark}(\nu)_n = V_{Tel}(\nu)_n + M_{Inst}(T_{Inst}, \nu)_n R_{Inst}(\nu). \qquad (3.1)$$

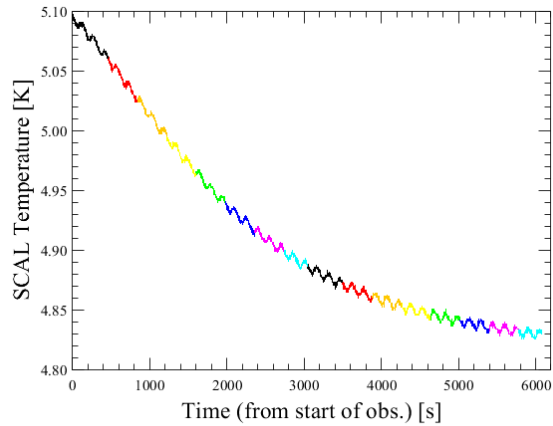

**Fig. 3**: **SPIRE Instrument temperature timeline, measured on the SCAL source, during a dark sky observation.** Each colour represents a group of five consecutive scans.

As mentioned above, the *Herschel* Telescope temperature can be assumed to be constant (within 10 mK) from scan to scan over the course of a single observation, and therefore the difference between any two spectra from scans i and j is

$$V_{Dark}(v)_i - V_{Dark}(v)_j = \left[ M_{Inst}(T_{Inst}, v)_i - M_{Inst}(T_{Inst}, v)_j \right] R_{Inst}(v). \qquad (3.2)$$

The instrument RSRF, $R_{Inst}(v)$, may be isolated by rearranging Eq. 3.2. Note that the convention that has been adopted is to always express the model terms as positive contributions and, in this case, this results in $R_{Inst}(v)$ being negative.

$$R_{Inst}(v) = \frac{V_{Dark}(v)_i - V_{Dark}(v)_j}{M_{Inst}(T_{Inst}, v)_i - M_{Inst}(T_{Inst}, v)_j} \qquad [\text{V GHz}^{-1} \, (\text{W m}^{-2} \, \text{Hz}^{-1} \, \text{sr}^{-1})^{-1}] \qquad (3.3)$$

In order to avoid small differences in instrument temperature in Eq. 3.3, one may split the observation into two halves – one half that contains scans where $n \geq N_{Scans}/2$, the other that contains scans where $n < N_{Scans}/2$ – the difference between scans in each half of an observation then becomes

$$V_{Dark}(v)_n - V_{Dark}(v)_{n - N_{Scans}/2} = R_{Inst}(v) \left[ M_{Inst}(T_{Inst}, v)_n - M_{Inst}(T_{Inst}, v)_{n - N_{Scans}/2} \right], \quad (3.4)$$

with the resulting instrument RSRF given by

$$R_{Inst}(v) = \frac{V_{Dark}(v)_n - V_{Dark}(v)_{n - N_{Scans}/2}}{M_{Inst}(T_{Inst}, v)_n - M_{Inst}(T_{Inst}, v)_{n - N_{Scans}/2}} \qquad [\text{V GHz}^{-1} \, (\text{W m}^{-2} \, \text{Hz}^{-1} \, \text{sr}^{-1})^{-1}] \qquad (3.5)$$

An example of the spectral half-differences and modelled SPIRE Instrument emission differences for one dark sky observation (OBSID=1342188673, OD=227, 26-12-2009) are shown in Fig. 4, and the resulting individual $R_{Inst}(v)$ curves for the observation are shown in Fig. 5.

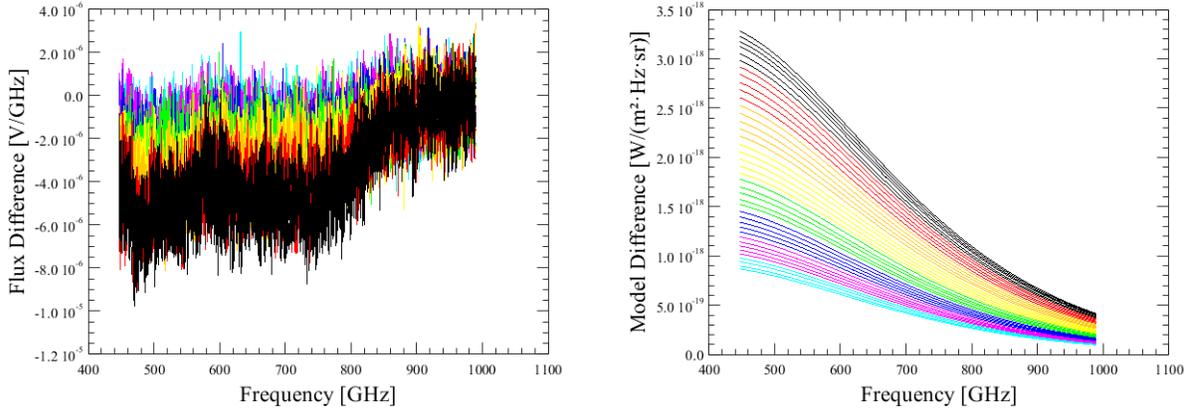

**Fig. 4**: **Example of spectral half-differences for detector SLWC3.** The different colours represent different scan combinations from the observation. The left plot shows the measured spectral differences in voltage density, and the right plot shows the differences in model intensity.

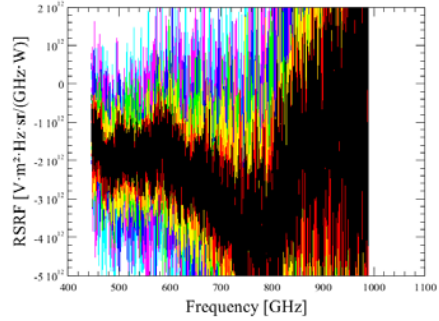

**Fig. 5**: **Empirical Derivation of the instrument RSRF**. OBSID=1342188673, detector=SLWC3, derived using half differences.

In principle, the measured spectra from every SPIRE FTS observation from any SPIRE FTS dark sky observation may be used to derive the instrument RSRF calibration curves given by Eq. 3.5. Indeed, if the stability of the telescope pointing is such that the spectral contribution from an astronomical target does not vary measurably over the course of an observation, then *all* SPIRE FTS observations may be used to derive $R_{Inst}(\nu)$. Moreover, since each $R_{Inst}(\nu)$ curve is normalized, a single $R_{Inst}(\nu)$ curve for each SPIRE FTS detector may be computed as the average of all of the individual $R_{Inst}(\nu)$ curves (see Eq. 3.6).

In order to minimize the variability in source radiation arising from pointing or other errors, only FTS observations of the dark sky were considered as candidates. In addition, noting from Eq. 3.5 that the individual elements that made up the average $R_{Inst}(\nu)$ curve are computed as a ratio of differences, a further restriction was enforced to maximize the temperature difference between each scan, thereby minimizing the overall noise of the average instrument RSRF. The choice of dark sky observations was therefore limited to those with a steeply changing instrument temperature profile similar to that shown in Fig. 3.

The instrument temperature profiles of all SPIRE FTS dark sky observations, the list of which may be found on the Herschel public TWiki page[1], were examined to determine the list of candidate observations for the empirical derivation of $R_{Inst}(\nu)$. Based on the aforementioned conditions, only six such candidates were identified, their instrument temperature profiles are displayed in Fig. 6 and details of which are given in Table 1.

| OBSID | Operational Day | $N_{Scans}$ | M1 [K] | M2 [K] | SPIRE Instrument [K] |
|---|---|---|---|---|---|
| 1342188195 | 217, 16-12-2009 | 100 | 88.062 ± 0.067 | 84.266 ± 0.012 | 4.797 ± 0.080 |
| 1342188673 | 227, 26-12-2009 | 80 | 88.271 ± 0.065 | 84.377 ± 0.024 | 4.915 ± 0.079 |
| 1342189120 | 240, 08-01-2010 | 80 | 88.062 ± 0.066 | 84.475 ± 0.021 | 4.793 ± 0.081 |
| 1342189541 | 250, 18-01-2010 | 100 | 88.605 ± 0.068 | 84.709 ± 0.027 | 5.048 ± 0.038 |
| 1342189892 | 261, 29-01-2010 | 100 | 88.727 ± 0.066 | 84.849 ± 0.021 | 4.857 ± 0.032 |
| 1342197456 | 383, 31-05-2010 | 100 | 88.196 ± 0.070 | 84.283 ± 0.024 | 4.786 ± 0.033 |

**Table 1: Dark Sky observations used to derive the original instrument RSRF ($R_{Inst}(\nu)$) calibration curves.**

---

[1] http://herschel.esac.esa.int/twiki/bin/view/Public/SpireDailyDarkObservations

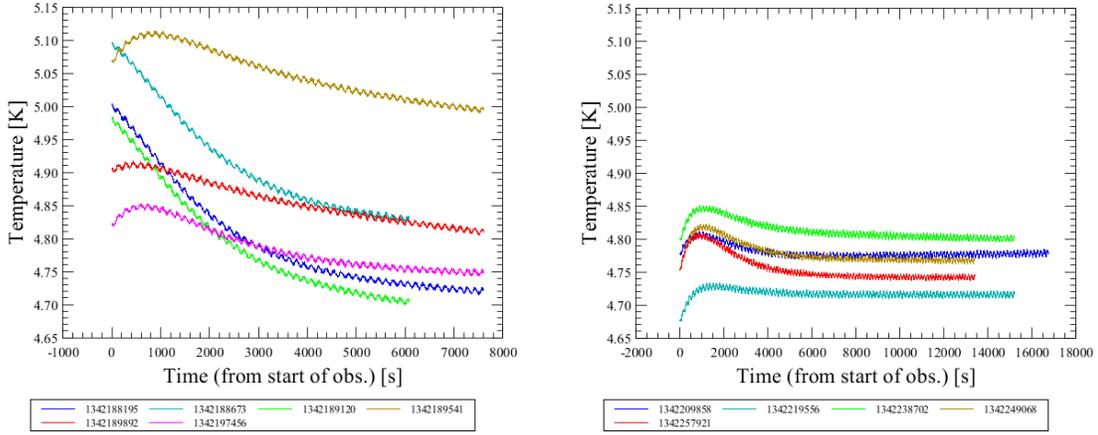

**Fig. 6: SPIRE Instrument temperatures.** The left panel shows the recorded values of the SPIRE Instrument thermometer for the dark sky observations originally used to derive the instrument RSRF calibration curves. The right panel shows the SPIRE Instrument temperatures for typical dark sky observations.

The $R_{Inst}(\nu)$ curves in units of V GHz$^{-1}$ (W m$^{-2}$ Hz$^{-1}$ sr$^{-1}$)$^{-1}$ for each detector were computed as the averages of all of the individual half-difference RSRF curves from each of the candidate dark sky observations as

$$\overline{R_{Inst}(\nu)} = \frac{1}{N_{Scans}/2} \sum_{n-N_{Scans}/2}^{N_{Scans}} R_{Inst}(\nu) = \frac{1}{N_{Scans}/2} \sum_{n-N_{Scans}/2}^{N_{Scans}} \left[ \frac{V_{Dark}(\nu)_n - V_{Dark}(\nu)_{n-N_{Scans}/2}}{M_{Inst}(T_{Inst}, \nu)_n - M_{Inst}(T_{Inst}, \nu)_{n-N_{Scans}/2}} \right] \quad (3.6)$$

The method used to derive the RSRF calibration curves described by Eq. 3.6 imposes a set of restrictions on the type of observations that can be used to derive $R_{Inst}(\nu)$, and therefore limits its final signal-to-noise ratio. In order to avoid such restrictions, thus enabling the use of all of the dark sky calibration observations that were made through the mission, a new method of combining the measured voltage density spectra of dark sky observations has been used from HIPE v11 onwards. Rather than focussing on the differences between individual spectra of the same observation, spectra from different observations, A and B, are compared,

$$V_{Meas}^A(\nu) = R_{Tel}(\nu)M_{Tel}(T_{Tel}^A, \nu) + R_{Inst}(\nu)M_{Inst}(T_{Inst}^A, \nu)$$
$$V_{Meas}^B(\nu) = R_{Tel}(\nu)M_{Tel}(T_{Tel}^B, \nu) + R_{Inst}(\nu)M_{Inst}(T_{Inst}^B, \nu). \quad (3.7)$$

Each of the spectra in Eq. 3.7 were divided by their respective telescope model, $M_{Tel}(T_{Tel}, \nu)$, yielding

$$V_{Meas}^A(\nu)/M_{Tel}(T_{Tel}^A, \nu) = R_{Tel}(\nu) + R_{Inst}(\nu)M_{Inst}(T_{Inst}^A, \nu)/M_{Tel}(T_{Tel}^A, \nu)$$
$$V_{Meas}^B(\nu)/M_{Tel}(T_{Tel}^B, \nu) = R_{Tel}(\nu) + R_{Inst}(\nu)M_{Inst}(T_{Inst}^B, \nu)/M_{Tel}(T_{Tel}^B, \nu). \quad (3.8)$$

The instrument response may be isolated as

$$R_{Inst}(\nu) = \frac{V_{Meas}^A(\nu)/M_{Tel}(T_{Tel}^A, \nu) - V_{Meas}^B(\nu)/M_{Tel}(T_{Tel}^B, \nu)}{[M_{Inst}(T_{Inst}^A, \nu)/M_{Tel}(T_{Tel}^A, \nu) - M_{Inst}(T_{Inst}^B, \nu)/M_{Tel}(T_{Tel}^B, \nu)]}. \text{ [V GHz}^{-1} \text{ (W m}^{-2} \text{ Hz}^{-1} \text{ sr}^{-1})^{-1}] \quad (3.9)$$

Note that Eq. 3.9 may be applied to a combination of any two voltage density spectra – i, j – from any two dark sky observations – A, B – for a given spectrometer detector. These derived instrument RSRF curves may then be averaged together to derive the final instrument response in units of V GHz$^{-1}$ (W m$^{-2}$ Hz$^{-1}$ sr$^{-1}$)$^{-1}$ for a given detector.

$$\overline{R_{Inst}(\nu)} = \frac{1}{N} \sum_{i,j,i \neq j}^{N} \frac{V_{Meas}^A(\nu)_i/M_{Tel}(T_{Tel}^A, \nu)_i - V_{Meas}^B(\nu)_j/M_{Tel}(T_{Tel}^B, \nu)_j}{[M_{Inst}(T_{Inst}^A, \nu)_i/M_{Tel}(T_{Tel}^A, \nu)_i - M_{Inst}(T_{Inst}^B, \nu)_j/M_{Tel}(T_{Tel}^B, \nu)_j]} \quad (3.10)$$

The resulting RSRFs derived from Eq. 3.10 are compared with the previous method (Eq. 3.6) in Section 5.

## 4. TELESCOPE CALIBRATION

After the application of the Instrument Correction step of the standard processing pipeline (Fulton et al, 2013), the measured spectrum of a dark sky observation contains no contribution from either an astronomical source or the SPIRE Instrument and therefore may be expressed as

$$V_{Inst-Corrected}(\nu) = V_{Tel}(\nu). \tag{4.1}$$

and the telescope RSRF for a single dark sky observation is given as

$$R_{Tel}(\nu) = \frac{V_{Inst-Corrected}(\nu)}{M_{Tel}(T_{Tel}, \nu)}. \tag{4.2}$$

Unlike the case for the original method for the derivation of the instrument RSRF described in Section 3, there were fewer restrictions for the derivation of the telescope RSRF. Contributions from all dark sky observations could be used to compute the overall telescope RSRF as

$$\overline{R_{Tel}(\nu)} = \sum_{n=1}^{N_{obs}} R_{Tel}^{Obs}(\nu) = \sum_{n=1}^{N_{obs}} \left[ \frac{V_{Inst-Corrected}^n(\nu)}{M_{Tel}^n(T_{Tel}, \nu)} \right] [V\ GHz^{-1}\ (W\ m^{-2}\ Hz^{-1}\ sr^{-1})^{-1}] \tag{4.3}$$

The only restriction on the telescope RSRF was related to the rest position of the SPIRE BSM. On OD 1011 (18-02-2012) the rest position of the SPIRE BSM for sparse sampled observations was changed by 1.7" (Swinyard et al, 2013). As the derivation of the telescope RSRF curves depends on the emission path from *Herschel* primary and secondary mirrors to the SPIRE detectors, it in turn depends on the rest position of the BSM (Fig. 1, Fig. 2). As a result, there were two epochs for the telescope RSRF curves: the first epoch based on and applicable to observations before OD 1011; the second epoch based on and applicable to observations on or after OD 1011.

When the method used to derive the instrument RSRF curves was updated for HIPE v11 (Section 3), a similar update was made to the method to derive the telescope RSRF. In a similar way to Eq. 3.8, each spectrum can be divided by its instrument model, $M_{Inst}(T_{Inst}, \nu)$, and the value of $R_{Tel}(\nu)$ isolated as

$$V_{Meas}^A(\nu)/M_{Inst}(T_{Inst}^A, \nu) = R_{Inst}(\nu) + R_{Tel}(\nu)M_{Tel}(T_{Tel}^A, \nu)/M_{Inst}(T_{Inst}^A, \nu)$$
$$V_{Meas}^B(\nu)/M_{Inst}(T_{Inst}^B, \nu) = R_{Inst}(\nu) + R_{Tel}(\nu)M_{Tel}(T_{Tel}^B, \nu)/M_{Inst}(T_{Inst}^B, \nu) \tag{4.4}$$

The final telescope RSRF in units of V GHz$^{-1}$ (W m$^{-1}$ Hz$^{-1}$ sr$^{-1}$)$^{-1}$ for a given detector is then given by

$$\overline{R_{Tel}(\nu)} = \frac{1}{N} \sum_{i,j,i\neq j}^N \frac{V_{Meas}(\nu)_i/M_{Inst}(T_{Inst}^A, \nu)_i - V_{Meas}(\nu)_j/M_{Inst}(T_{Inst}^B, \nu)_j}{\left[ M_{Tel}(T_{Tel}^A, \nu)_i/M_{Inst}(T_{Inst}^A, \nu)_i - M_{Tel}(T_{Tel}^B, \nu)_j/M_{Inst}(T_{Inst}^B, \nu)_j \right]} \tag{4.5}$$

## 5. COMPARISONS AND RESULTS

Sections 3 and 4 describe the method used to derive the instrument and telescope RSRFs, and the changes made for HIPE v11. As noted in Section 3, there are fewer restrictions on the data that can be used to derive the RSRFs, particularly for the instrument RSRF, using the new derivation procedure. The updated derivation procedure only requires that the instrument temperatures of $V_{Dark}^A(\nu)_i$ and $V_{Dark}^B(\nu)_j$ be different, and so many more dark sky spectral combinations could therefore be included in the derivation as shown in Table 2.

| Number of Spectral | Original Calibration Method | | New Calibration Method | |
|---|---|---|---|---|
| Combinations | BSM1 | BSM2 | BSM1 | BSM2 |
| $R_{Inst}(\nu)$ | 140 | 140 | 1,700,078 | 2,936,350 |
| $R_{Tel}(\nu)$ | 2,242 | 1,200 | 1,700,078 | 2,936,350 |

**Table 2: Number of spectral combinations used to derive the RSRFs.** BSM 1 and BSM 2 refer to the rest position of the SPIRE Beam Steering Mirror that was changed on OD1011 (18-02-2012).

The reduction in noise due to the increased number of spectra used to derive the RSRFs is evident from a comparison between the initial and the updated calibration curves (Fig. 7, Fig. 8) and their statistical errors (Fig. 9), the latter showing a decrease in the statistical errors of two orders of magnitude.

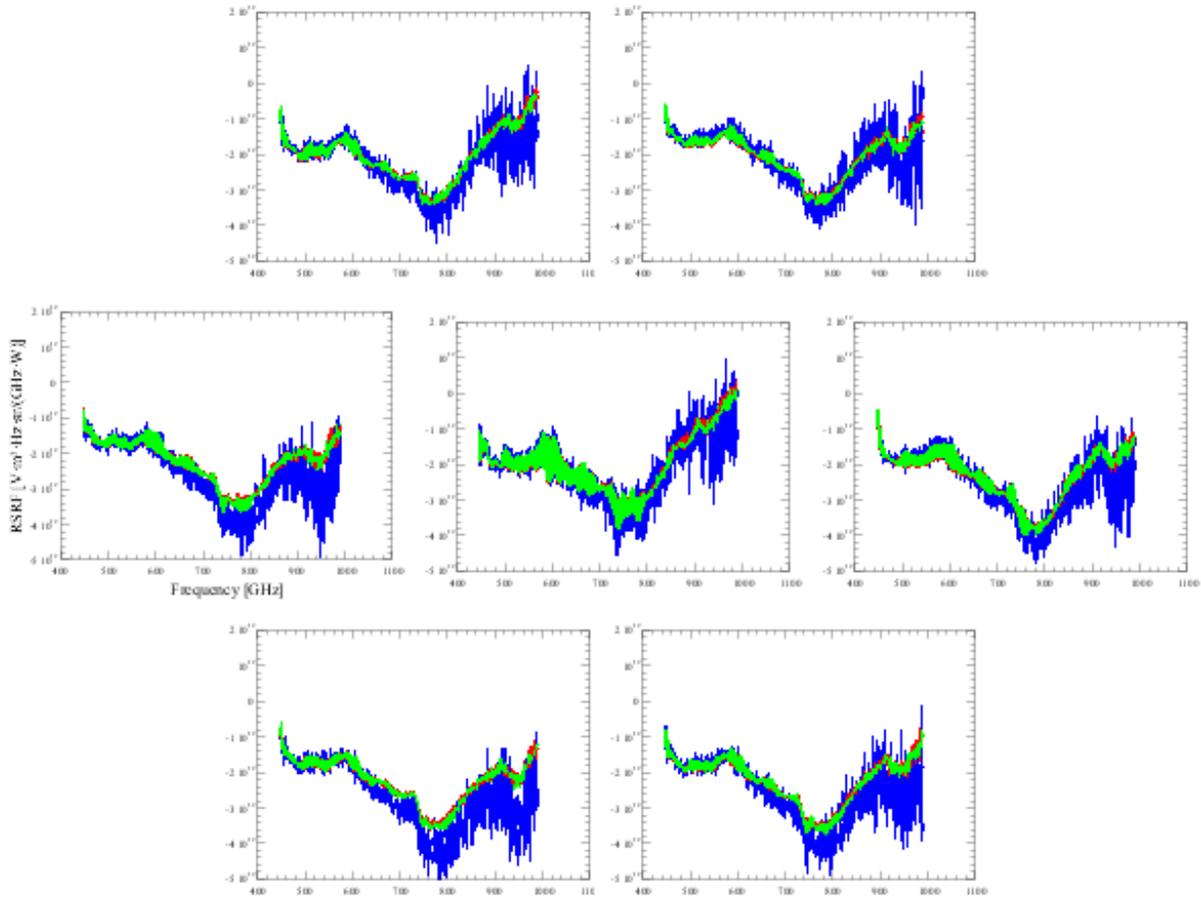

**Fig. 7**: **Instrument RSRFs: All Unvignetted SLW detectors.** The blue curves represent the original calibration method; the red and green curves represent the new calibration method for the epochs before and after the change of the BSM home position, respectively.

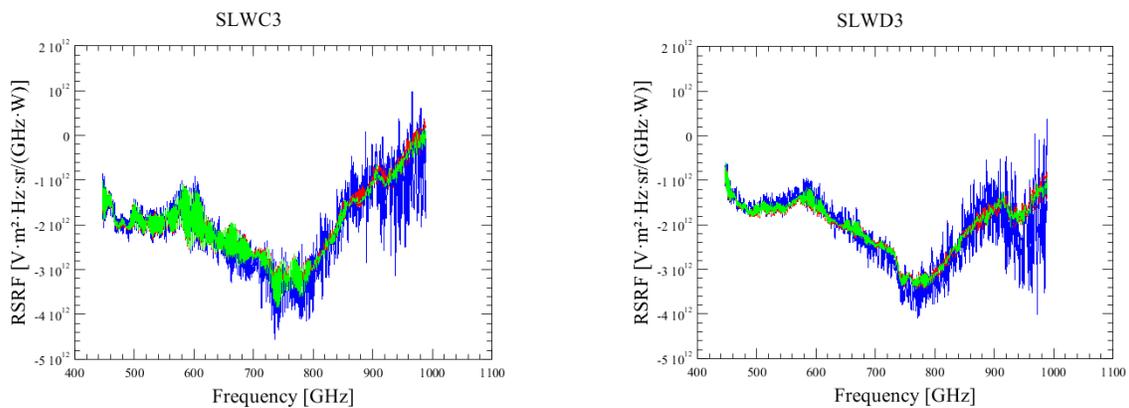

**Fig. 8**: **Instrument RSRFs: left, detector SLWC3; right, detector SLWD3.** The blue curves represent the original calibration method; the red and green curves represent the new calibration method for the epochs before and after the change of the BSM home position, respectively.

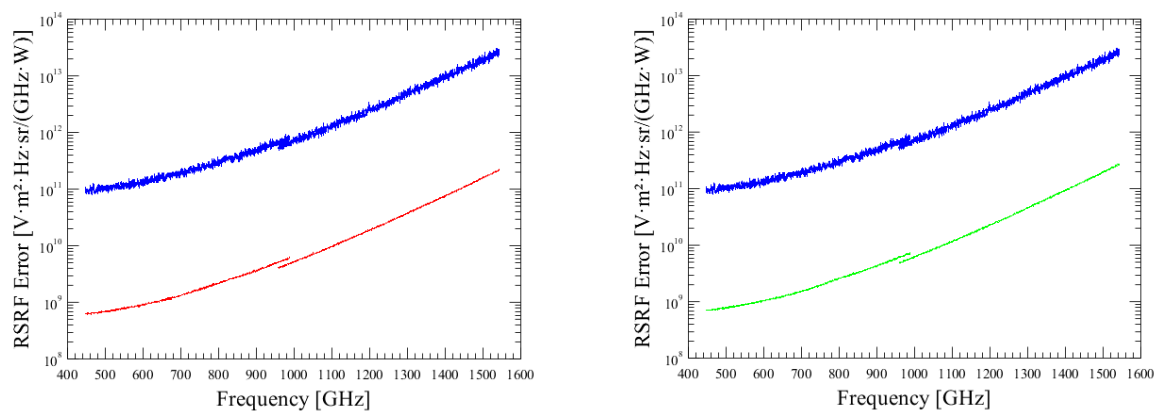

**Fig. 9**: **Statistical Errors, Instrument RSRFs for detectors SLWC3 and SSWD4: left, BSM position before OD1011 (18-02-2012); right, BSM position after OD1011 (18-02-2012).** The blue curves represent the original calibration method; the red and green curves represent the new calibration method for the epochs before and after the change of the BSM home position, respectively.

Similarly, the telescope RSRFs and RSRF errors for the central detectors of the two spectrometer arrays are shown in Fig. 10 and Fig. 11. The comparison of the statistical errors reveals a factor of 2 gain for the updated calibration method.

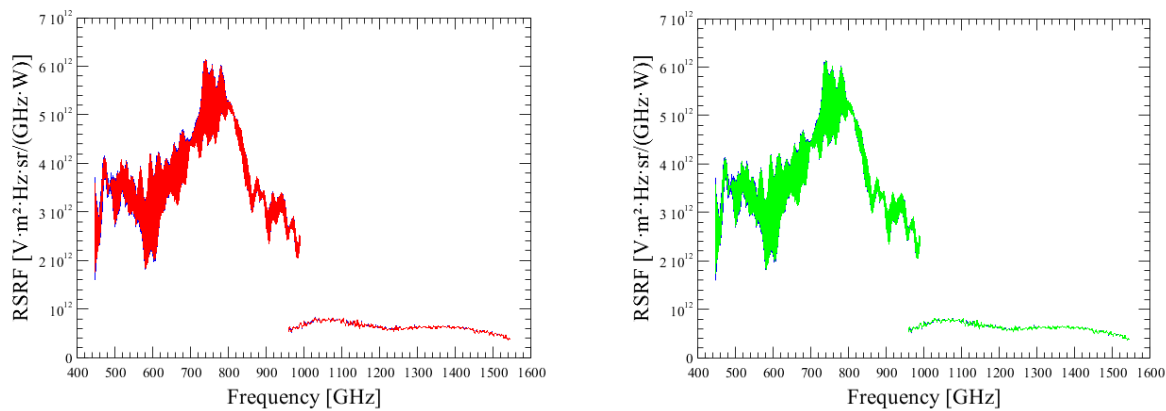

**Fig. 10**: **Telescope RSRFs, detectors SLWC3 and SSWD4: left, BSM position before OD1011 (18-02-2012); right, BSM position after OD1011 (18-02-2012).** The blue curves represent the original calibration method; the red and green curves represent the new calibration method for the epochs before and after the change of the BSM home position, respectively.

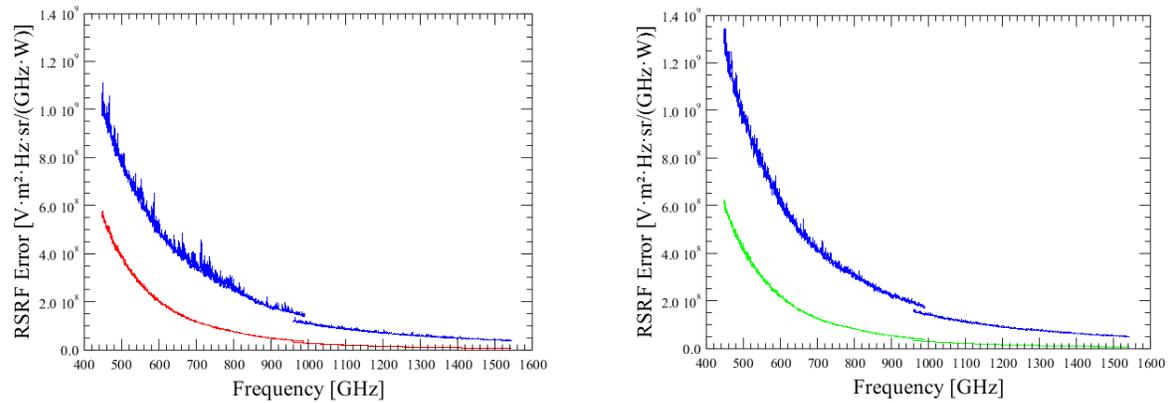

**Fig. 11**: **Statistical Errors, Telescope RSRFs, detectors SLWC3 and SSWD4: left, BSM position before OD1011 (18-02-2012); right, BSM position after OD1011 (18-02-2012).** The blue curves represent the original calibration method; the red and green curves represent the new calibration method for the epochs before and after the change of the BSM home position, respectively.

The statistical errors provide a measure of how the instrument and telescope RSRFs contribute to the overall error budget of a calibrated SPIRE FTS spectrum. The plots shown in Fig. 12 show how the change in the methods used to derive the RSRFs have had an impact on the noise of a calibrated SPIRE FTS spectrum. The 1-σ noise is computed by first subtracting the continuum from the spectrum by way of a polynomial fit and then computing the standard deviation over a series 50 GHz bins. As can be seen, for spectra that are free of line contributions, the noise is reduced by a factor of 2-3 throughout the spectral passbands with larger gains realized at the low frequency end of SLW. Similar gains are also noted in that region for the spectra of sources with lines.

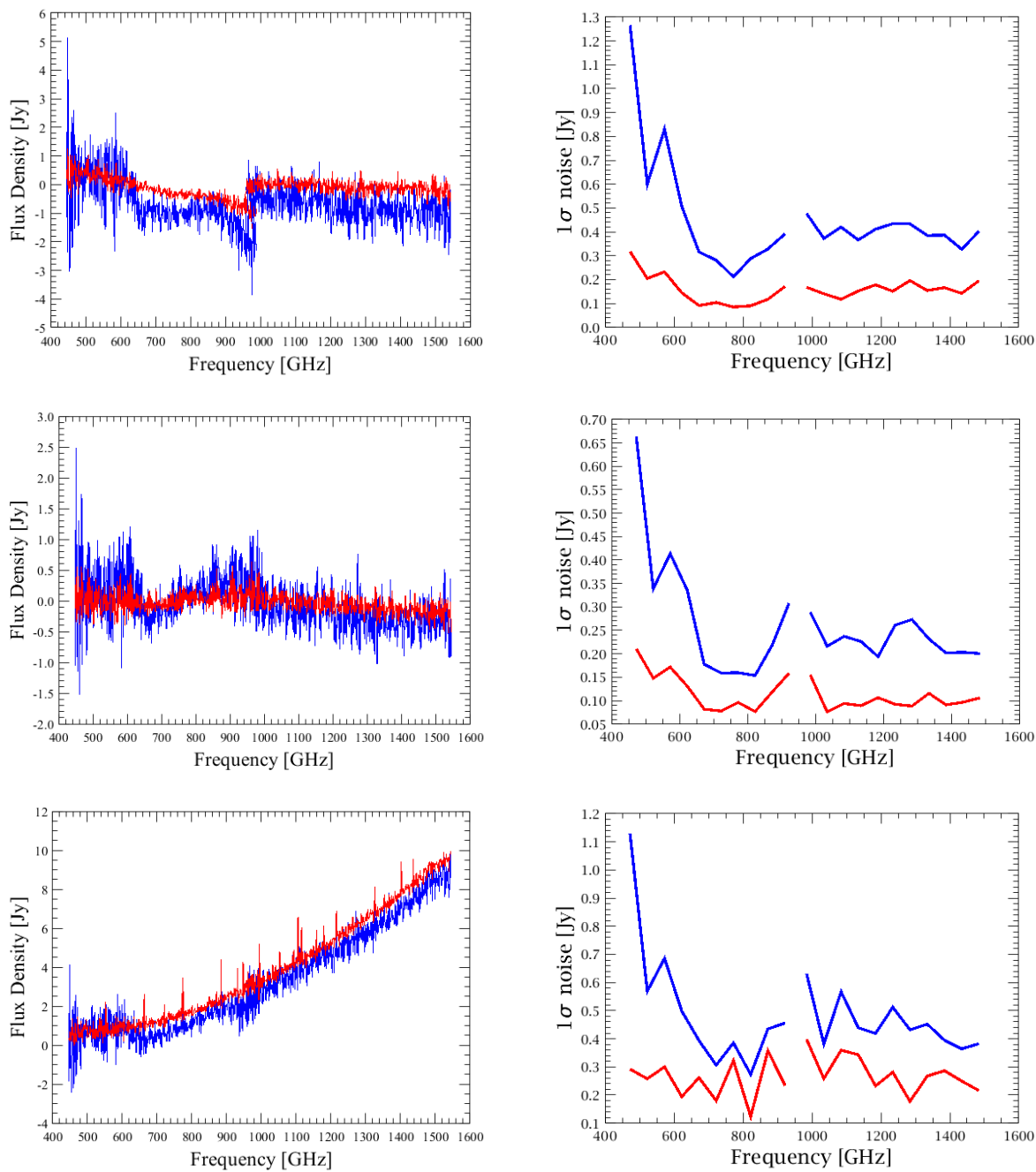

**Fig. 12: Calibrated Spectra and noise for selected SPIRE FTS calibration and science observations.** The blue curves represent spectra computed using with RSRFs derived the original method; the red curves represent spectra using with RSRFs derived the new method.

The noise of a specific SPIRE FTS spectrum can be normalized to account for the duration of the observation. In this way, an overall measurement of the sensitivity can be computed as a combination of the sensitivity of individual observations. Furthermore, the overall sensitivity can be compared to the values from the SPIRE Observer's Manual (2011) that were used for observation proposal purposes (see Fig. 13). Note that only dark sky observations were included in the computation of the overall sensitivity.

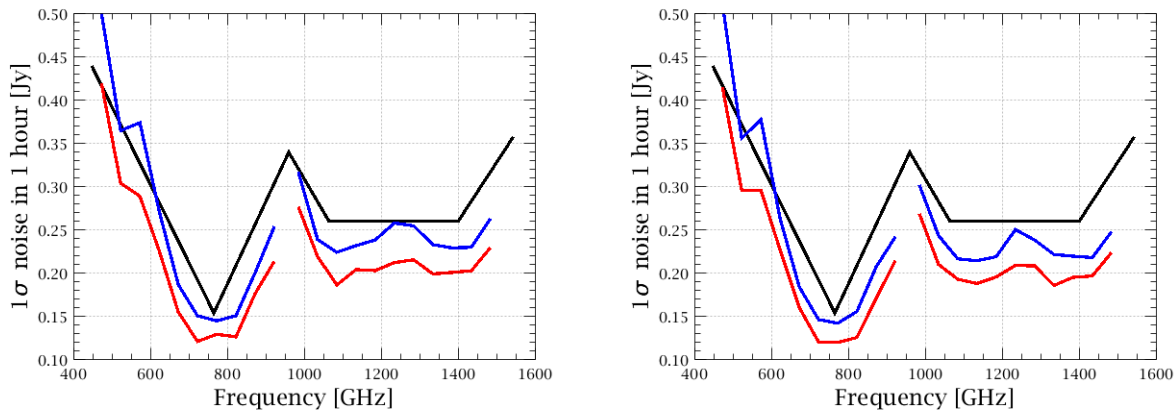

**Fig. 13**: **Sensitivity for sparse mode, high spectral resolution SPIRE FTS observations: left, BSM position before OD1011 (18-02-2012); right, BSM position after OD1011 (18-02-2012).** The black curves are those values listed in the SPIRE Observer's Manual; the blue curves represent the sensitivity derived with the original RSRFs; the red curves represent the sensitivity derived with the new RSRFs.

As the curves in Fig. 13 show, while the SPIRE FTS calibrated spectra were achieving or exceeding the predicted sensitivity with the original RSRF calibration curves, a further gain of 23% for SLW and 21% for SSW has been achieved with the new calibration curves.

## 6. CONCLUSION

An update to the method used to derive the Relative Spectral Response Functions for high spectral resolution, sparse sampling SPIRE Fourier Transform Spectrometer observations has been presented. The updated derivation method was necessary in order to utilize a larger sample of the SPIRE FTS calibration data and in turn reduce the noise imparted to the final calibrated spectra. A comparison of the calibration curves themselves showed that the instrument RSRF contribution to the overall error budget has been reduced by a factor of 100, while that for the telescope RSRF has been reduced by a factor of 2. The improved calibration quality results in a reduction in the noise in the calibrated spectra by factors of 2-3, depending on the nature of the observation. A comparison of the normalized sensitivity reveals that an overall gain of 23% for SLW and of 21% for SSW has been achieved with the new calibration curves in HIPE v11. This calibration is available in pipeline produced data from HIPE v11 onwards.

## ACKNOWLEDGEMENTS

*Herschel* is an ESA space observatory with science instruments provided by European-led Principal Investigator consortia with important participation from NASA. SPIRE has been developed by a consortium of institutes led by Cardiff University (UK) and including Univ. Lethbridge (Canada); NAOC (China); CEA, LAM (France); IFSI, Univ. Padua (Italy); IAC (Spain); Stockholm Observatory (Sweden); Imperial College London, RAL, UCL-MSSL, UKATC, Univ. Sussex (UK); and Caltech, JPL, NHSC, Univ. Colorado (USA). This development has been supported by national funding agencies: CSA (Canada); NAOC (China); CEA, CNES, CNRS (France); ASI (Italy); MCINN (Spain); SNSB (Sweden); STFC (UK); and NASA (USA).